\documentstyle[aps,epsfig,twocolumn]{revtex}
\begin{document}
\draft

\title{$\eta-\eta'$-Glueball mixing from photon-meson transition form
factors and decay ratio $D_s\to\eta l\nu/\eta'l\nu$}
\author{V.V.Anisovich\thanks{e-mail: anisovic@lnpi.spb.su}}
\address{St.Petersburg Nuclear Physics Institute, Gatchina, 188350,
Russia}
\author{D.V.Bugg}
\address{Queen Mary and Westfield College, Mile End Rd., London E1
4NS, UK}
\author{D.I.Melikhov\thanks{e-mail: melikhov@monet.npi.msu.su}}
\address{Nuclear Physics Institute, Moscow State University, 119899,
Moscow, Russia}
\author{V.A.Nikonov\thanks{e-mail: nikon@rec03.pnpi.spb.ru}}
\address{St.Petersburg Nuclear Physics Institute, Gatchina, 188350,
Russia}
\maketitle
\widetext

\begin{abstract}
We have determined the $\eta/\eta'$ mixing angle and the regions of
the allowed admixture of the glueball component with $J^{PC}=0^{-+}$
in $\eta$ and $\eta'$, based on transition form factors
$\eta\to\gamma\gamma^*$ and $\eta'\to\gamma\gamma^*$ at
$0\le Q^2\le20\mbox{ GeV}^2\mbox{/c}^2$, and the branching ratio
$D_s\to\eta l\nu/\eta'l\nu$. For $\eta$ and $\eta'$ wave functions,
$\eta=\cos\alpha\;(\cos\theta\;n\bar n-\sin\theta\;s\bar s)
+\sin\alpha\;G$ and
$\eta'=\cos\alpha'\;(\sin\theta'\;n\bar n+\cos\theta'\;s\bar s)
+\sin\alpha'\;G$
where $n\bar n=(u\bar u+d\bar d)/\sqrt{2}$ and $G$ means a glueball, a
shape of the allowed region in the $(\theta,\alpha,\alpha')$-space is
determined which is located inside the borders
$0.56\le\sin\theta\le0.66$, $\sin^2\alpha\le0.08$,
$\sin^2\alpha'\le0.12$. This corresponds to
$\theta_{\mathrm{P}}=-17.0^\circ\pm2.6^\circ$.
\end{abstract}

\narrowtext

\vspace{1cm}
The problem of a precise determination of the $\eta$ and $\eta'$
contents lies in the fact that these states have been considered for a
long time as candidates for states with a significant admixture of
glueball components. The actual presence of such mixing can be
determined by analyzing extensive experimental information. The
conventional way of determining the $\eta/\eta'$ mixing angle from
mass formulae suggests values in the range
$-22^\circ<\theta_{\mathrm{P}}<-10^\circ$ ($\theta_{\mathrm{P}}$ is
the $\eta_1/\eta_8$ mixing angle
$\theta=54.74^\circ+\theta_{\mathrm{P}}$), depending on whether linear
or quadratic mass formulae are used \cite{datagr}. Experiments on
meson transitions which involve $\eta$ and $\eta'$ open a possibility
of an alternative determination of the pseudoscalar mixing angle from
form factor physics.

We consider two different processes for a determination of the mixing
angles: (i) transition form factors $\eta,\eta'\to\gamma\gamma^*$ at
$Q^2\le20\mbox{ GeV}^2\mbox{/c}^2$, including $\eta$ and $\eta'$
partial decay widths into $\gamma\gamma$ which correspond to $Q^2=0$,
and (ii) semileptonic electroweak transitions $D_s\to\eta
l\nu/\eta'l\nu$.

\section{Transitions $\eta\to\gamma\gamma^*$ and
$\eta'\to\gamma\gamma^*$}

The approach developed in Refs.\ \cite{amn,amn2} to the description of
elastic and transition form factors lies in taking into account a
truly Strong-QCD\footnote[1]{The notation "Strong-QCD" instead of
"nonperturbative QCD" is suggested by F.E. Close \cite{close}.}
part and the $O(\alpha_s)$\\
$$\vspace{1.65cm}$$ 
\noindent          
corrections to the form factors; thus we
can expect to give a reasonable description in the region of
intermediate momentum transfers. Our strategy is as follows. The
accurate structure of the soft pion wave function is determined by
fitting the data on elastic pion form factor \cite{amn}. Then, having
determined the pion wave function, we extract the soft photon wave
function from data on the $\pi^0\to\gamma\gamma^*$ transition form
factor; we find it to be pretty similar to the pion soft wave function
\cite{amn2}, quite in line with vector meson dominance. Furthermore,
assuming universality of the wave functions of the ground-state
pseudoscalar meson nonet, we have at hand the nonstrange component of
the $\eta$ and $\eta'$ wave function. The main SU(3) breaking effect
in the strange and nonstrange components of the meson wave functions
comes from the mass difference of the nonstrange and strange quarks;
hence we can also determine the wave function of the strange
component. We thus calculate the $\eta,\eta'\to\gamma\gamma^*$
transition form factors \cite{amn2} and can analyze them by comparison
with the data. All lengthy technical details for the calculation of
the elastic and transition form factors have been given in Refs.\
\cite{amn,amn2} and will not be presented here. We only briefly
outline the main steps.

We split the meson wave function into soft and hard components,
$\Psi^{\mathrm{S}}_\pi$ and $\Psi^{\mathrm{H}}_\pi$ such that
$\Psi^{\mathrm{S}}$ is large at $s=(m^2+k_\perp^2)/(x(1-x))<s_0$ where
$s$ is the $q\bar q$ invariant energy squared and $m$ is quark mass;
$\Psi^{\mathrm{H}}$ prevails at $s>s_0$. The parameter $s_0$ is a
boundary of the soft and hard regions and is expected to have the
value of several GeV$^2$. We perform the splitting of the wave
function into the soft and the hard components using a simple
step-function ansatz
\begin{equation}\label{psidec}
\Psi_\pi=\Psi^{\mathrm{S}}_\pi\ \theta(s_0-s)+\Psi^{\mathrm{H}}_\pi\
\theta(s-s_0)\ .
\end{equation}

The hard component $\Psi^{\mathrm{H}}_\pi$ is represented as a
convolution of the one-gluon exchange kernel $V^{\alpha_s}$ with
$\Psi^{\mathrm{S}}_\pi$
\begin{equation}\label{psihard}
\Psi^{\mathrm{H}}_\pi=V^{\alpha_s}\otimes\Psi^{\mathrm{S}}_\pi\ ,
\end{equation}
thus one comes to the following expansion of the elastic pion form
factor as a series in $\alpha_s$:
\begin{equation} \label{ffdec}
F_\pi=F^{\mathrm{SS}}_\pi+2F^{\mathrm{SH}}_\pi+O(\alpha^2_s)\ ,
\end{equation}
where $F^{\mathrm{SS}}_\pi$ is a truly Strong-QCD part of the form
factor, and $F^{\mathrm{SH}}_\pi$ is an $O(\alpha_s)$ term with
one-gluon exchange. The first term dominates the pion form factor at
small $Q^2$. The second term gives a minor contribution at small
$Q^2$ but provides the leading $\alpha_s(Q^2)/Q^2$ behaviour of the
elastic form factor at large $Q^2$.

The soft wave function $\Psi^{\mathrm{S}}_\pi$ is responsible for the
meson elastic form factor behaviour both at small and moderately large
$Q^2$. The value of $s_0$ and the soft wave function are variational
parameters of this approach.

\begin{figure}
\centerline{\epsfig{file=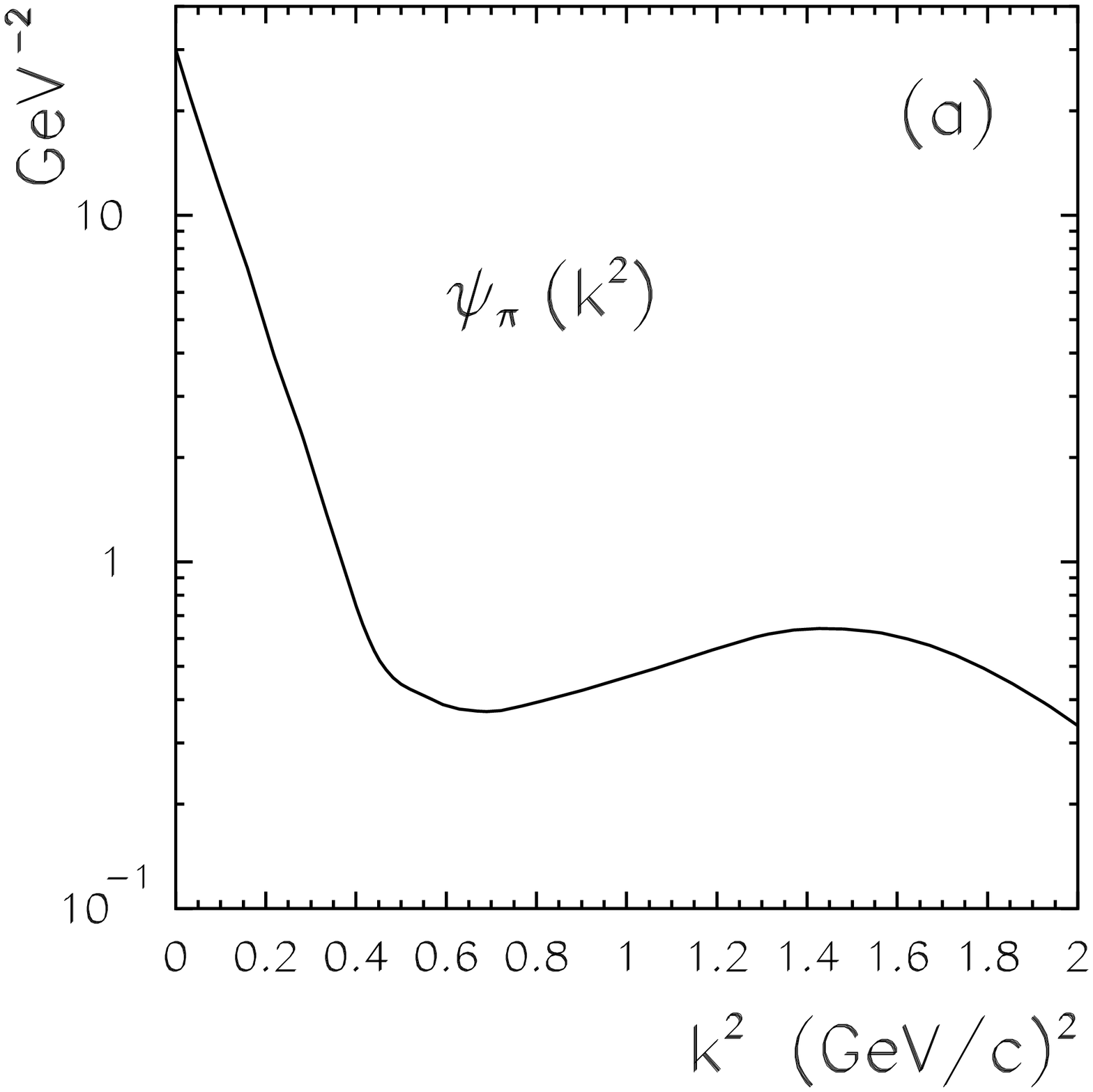, width=4cm}
            \epsfig{file=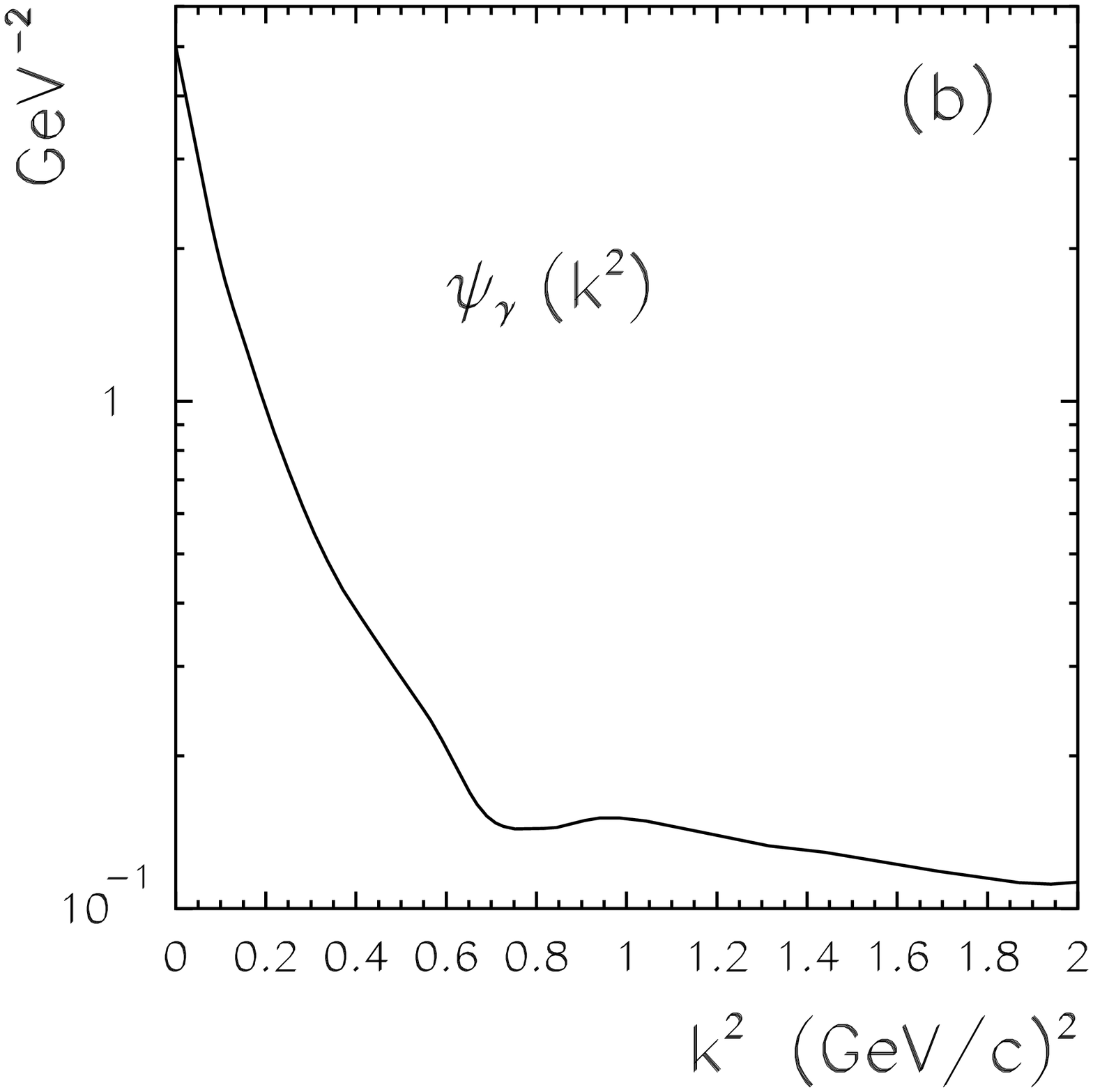,width=4cm}}
\caption{The reconstructed wave functions of pion (a) and photon (b).}
\label{fig:wfs} \end{figure}

\begin{figure}
\centerline{\epsfig{file=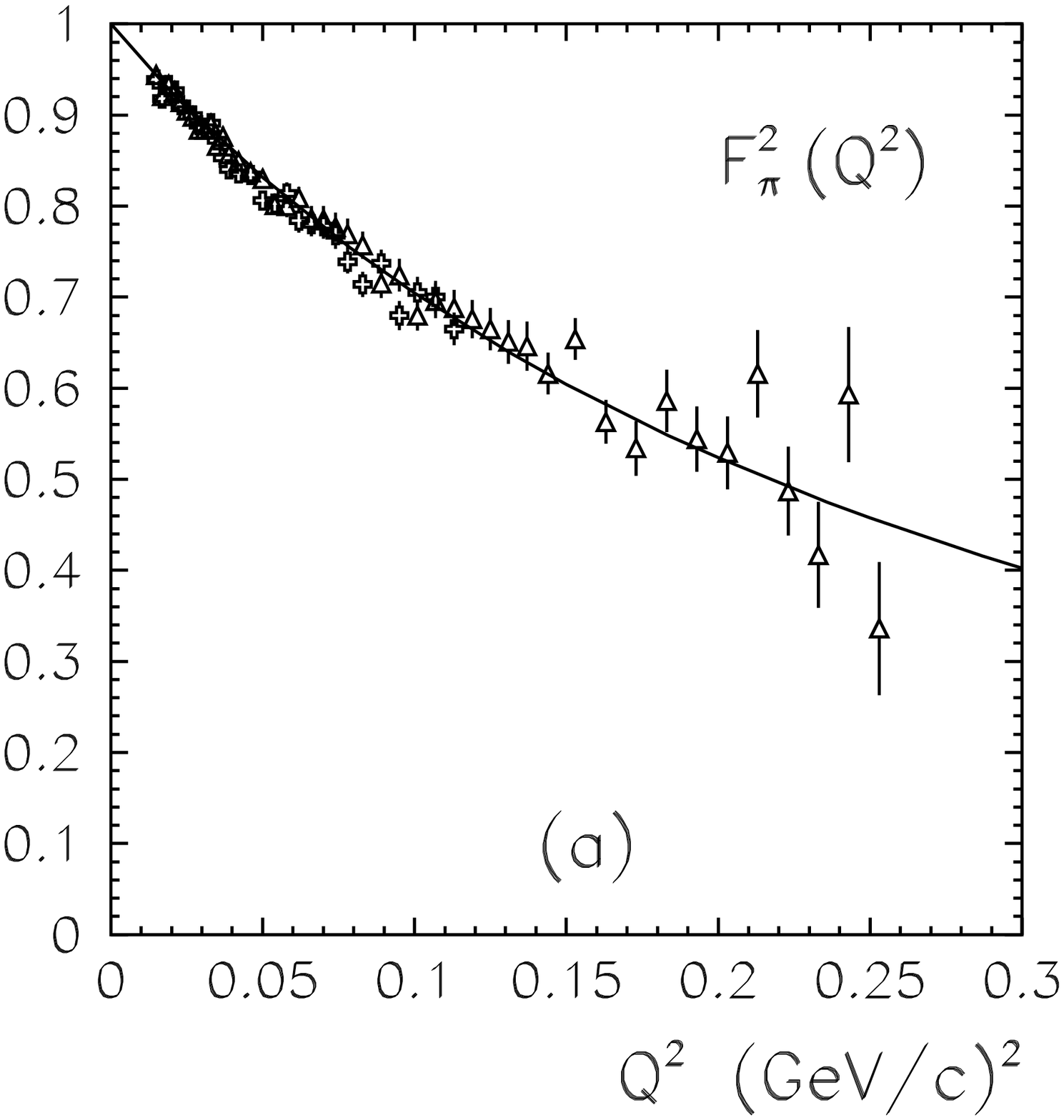,   width=3cm}
            \epsfig{file=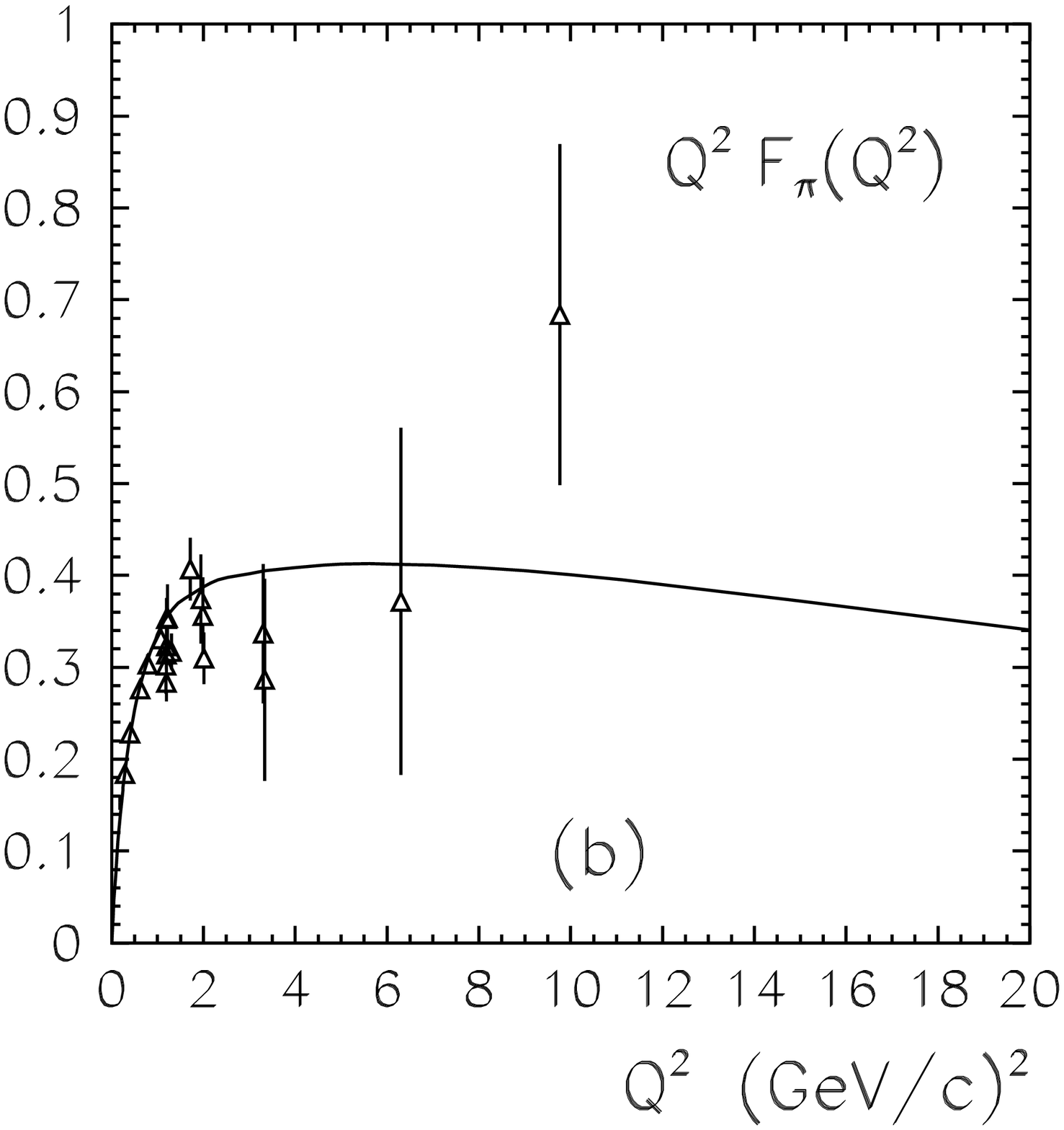, width=3cm}
            \epsfig{file=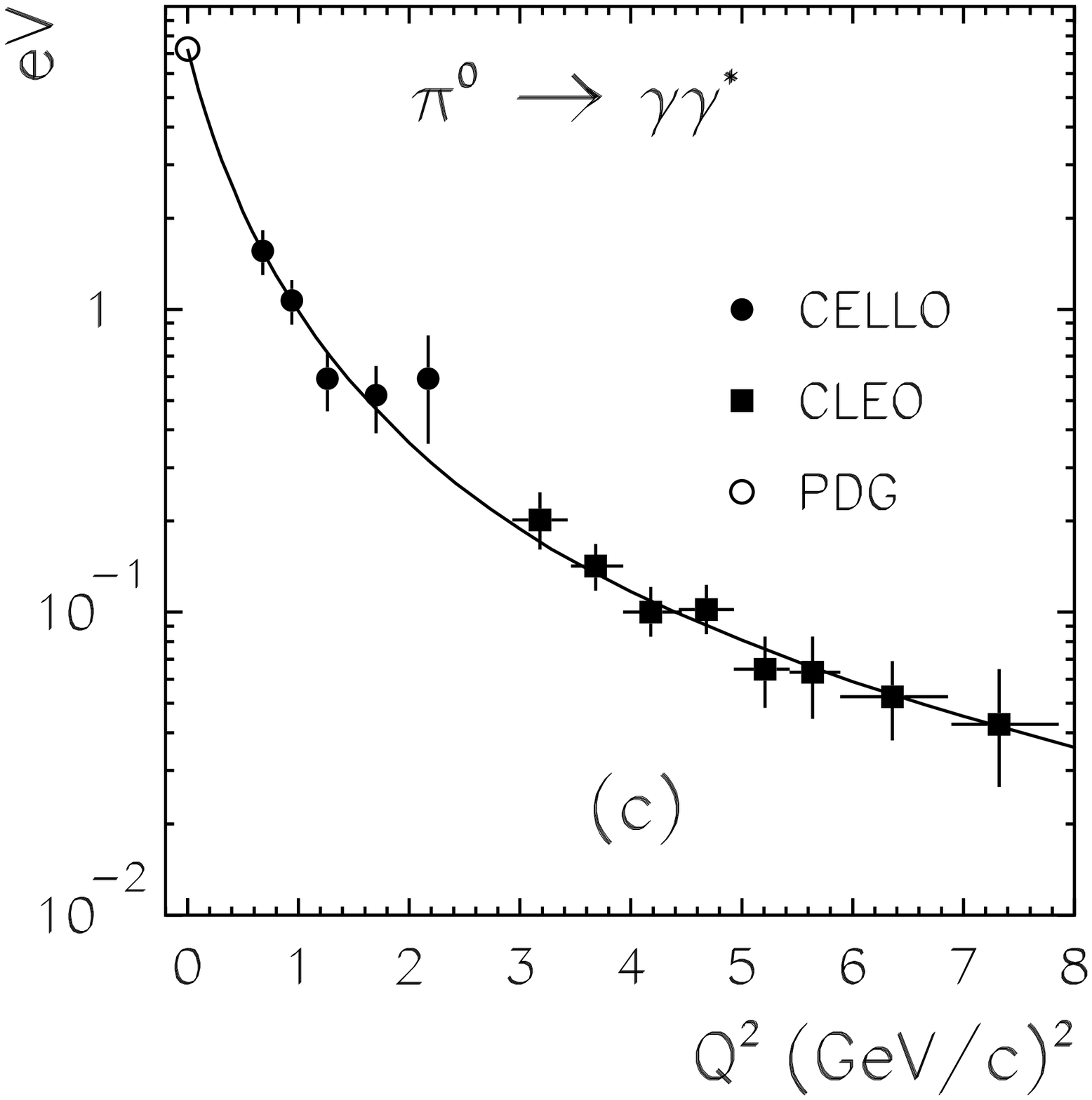,width=3cm}}
\caption{a,b) Description of pion form factor with the pion wave
function of Fig.\ \ref{fig:wfs}a;
c) The quantity $\Gamma_{\gamma\pi^0}(Q^2)=$
$\frac\pi4\alpha^2m_\pi^3 F^2_{\gamma\pi^0}(Q^2)$ and its description
by the diagrams of Fig.~\ref{fig:diagr} with photon wave function
presented in Fig.\ \ref{fig:wfs}b.}
\label{fig:fits}\end{figure}

Applying this strategy to the pion elastic form factor numerical
analysis we have found $s_0=9\mbox{ GeV} ^2$ provides the best
description of the data. This value corresponds to an extended soft
region and thus we relate a large portion of the pion form factor to
the soft contribution. Although particular values of the soft and hard
contributions to the form factor are model-dependent quantities we
find that a good description of the form factor at small $Q^2$ yields
a substantial soft contribution to the form factor at $Q^2\simeq
10-20\mbox{ GeV} ^2$. It is convenient to represent
$\Psi^{\mathrm{S}}_\pi$ in terms of the relative momentum
$\vec k^2=(s-4m^2)/4$ GeV with $m=0.35$ GeV as follows:
\begin{equation}
\Psi^{\mathrm{S}}_\pi=\psi_\pi(\vec k^2)=\frac{g_\pi(\vec k^2)}
{\vec k^2 +\kappa_0^2}\ ,
\end{equation}
where $\kappa_0^2=0.1176\mbox{ GeV}^2$ \cite{amn}. The
reconstructed wave function $\psi_\pi(\vec k^2)$ is shown in Fig.\
\ref{fig:wfs}a while Figs.\ \ref{fig:fits}a,b give the elastic
pion form factor calculated with this wave function (experimental
data from Ref.\ \cite{pipiexp}).

Similarly, for the description of the photon-pion transition form
factor we introduced the photon $q\bar q$ wave function and split it
into soft and hard components as follows
\begin{equation} \label{psigam}
\Psi_\gamma=\Psi^{\mathrm{S}}_\gamma\ \theta(s_0-s)
+\Psi^{\mathrm{H}}_\gamma\ .
\end{equation}

The soft component describes a hadronic $q\bar q$ structure of the
soft photon just in the spirit of vector meson dominance and can be
expected to have the same structure as the soft wave function of a
meson. However, the hard component of the photon wave function has an
important distinction compared with the hadron case: in addition to
the perturbative tail of the soft part of the wave function Eq.\
(\ref{psihard}), the hard component of the photon wave function
contains also a standard pointlike QED $q\bar q$-component such that
\begin{equation}\label{psihardg}
\Psi_\gamma^{\mathrm{H}}=V^{\alpha_s}\otimes\Psi_\gamma^{\mathrm{S}}
+\Psi_\gamma^{\mathrm{Pt}}\ .
\end{equation}

\begin{figure}
\centerline{\epsfig{file=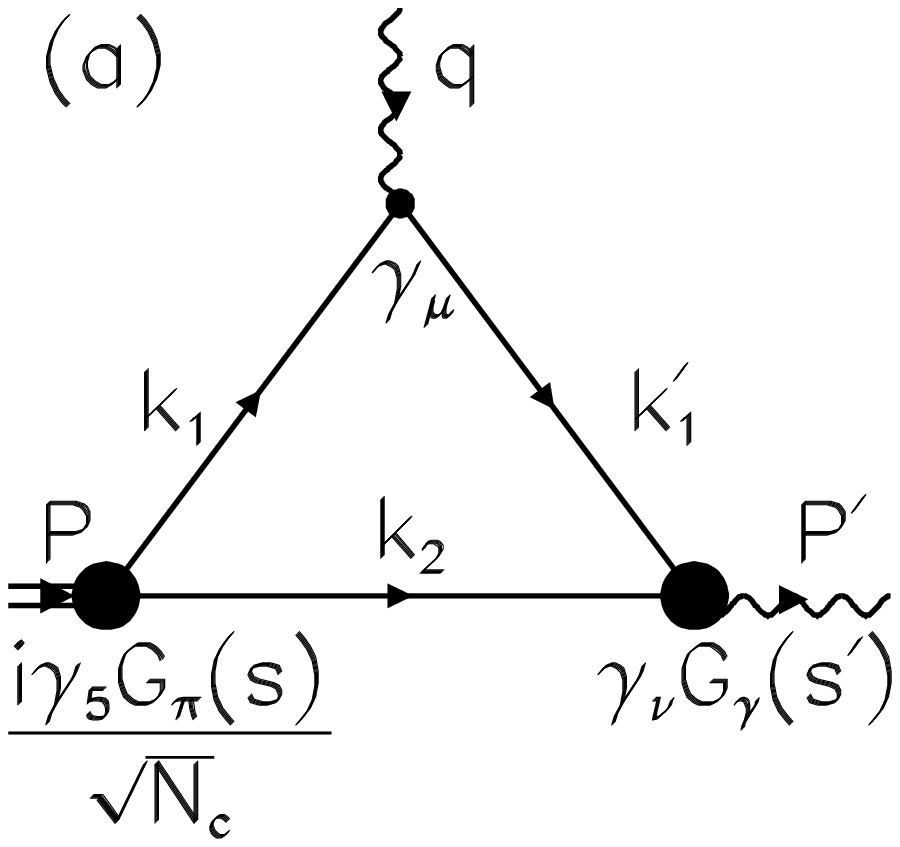, width=3cm}
            \epsfig{file=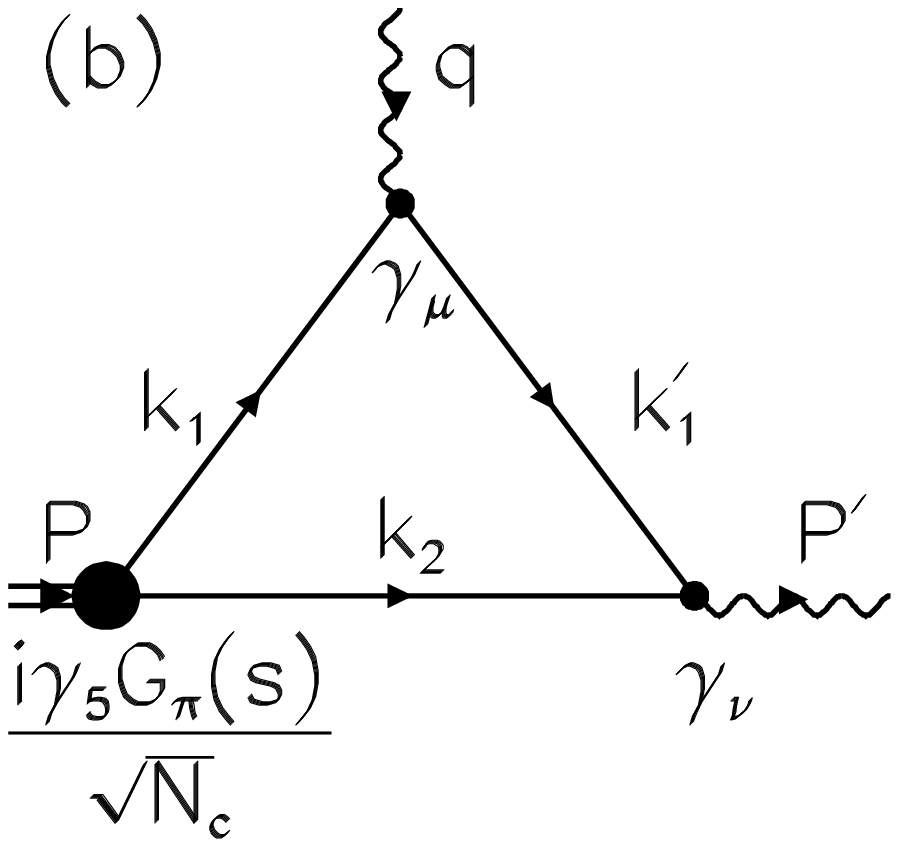,width=3cm}}
\centerline{\epsfig{file=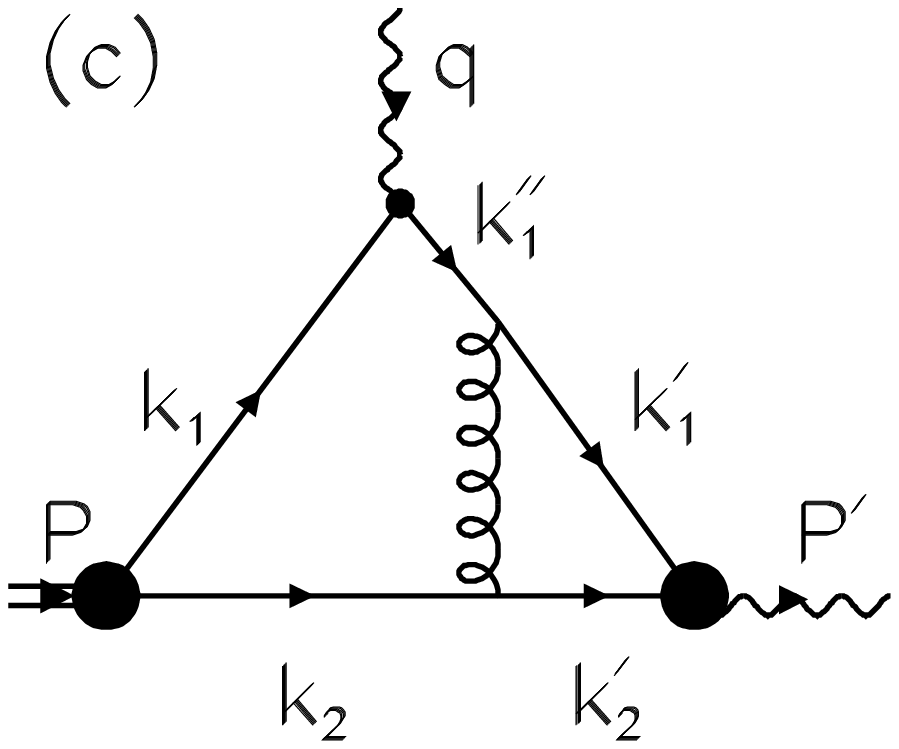,width=3cm}
            \epsfig{file=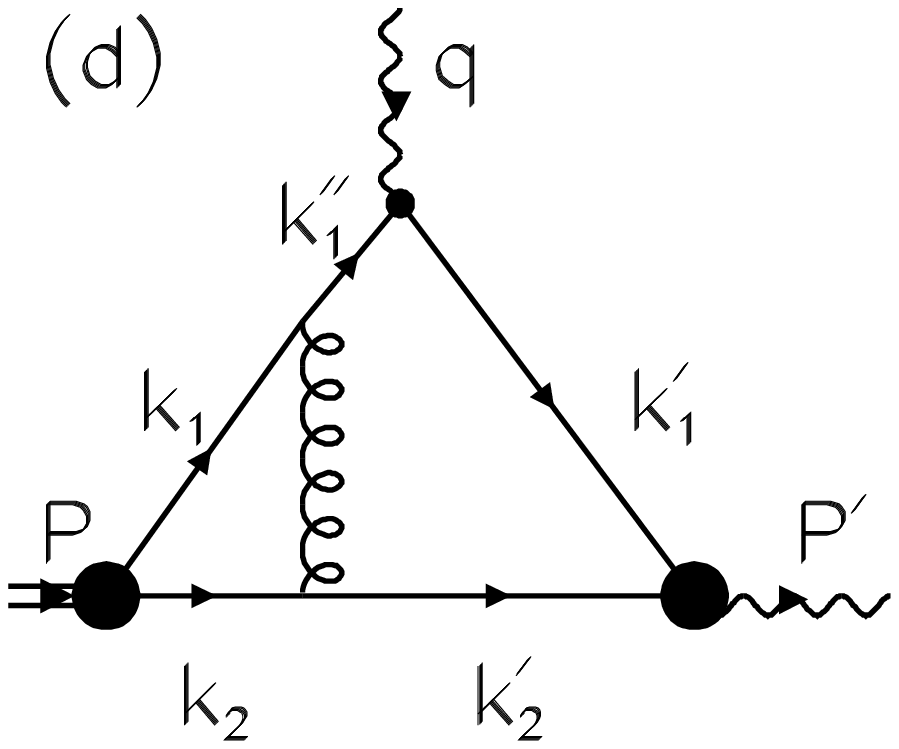,width=3cm}}
\caption{Diagrams relevant to the description of the meson-photon
transition form factor at low and moderately high $Q^2$:
$F^{\mathrm{SS}}$-term (a), $F^{\mathrm{SPt}}$-term (b),
$F^{\mathrm{SH(1)}}$-, $F^{\mathrm{SH(2)}}$-terms (c,d).}
\label{fig:diagr}\end{figure}

The corresponding expansion of the photon-meson transition form factor
has the form (see Fig.\ \ref{fig:diagr}):
\begin{equation}\label{ffdecg}
F_{\gamma\pi^0}= F^{\mathrm{SS}}+F^{\mathrm{SPt}}+F^{\mathrm{SH(1)}}
+F^{\mathrm{SH(2)}}\ ,
\end{equation}
where $F^{\mathrm{SH(1)}}+F^{\mathrm{SH(2)}}$ are $O(\alpha_s)$ terms.

At small $Q^2$, the $F^{\mathrm{SS}}$-part dominates the transition
form factor. At large $Q^2$, the soft-pointlike term
$F^{\mathrm{SPt}}$ gives the leading $1/Q^2$ falloff whereas the
contribution of the $O(\alpha_s)$ terms,
$F^{\mathrm{SH(1)}}+F^{\mathrm{SH(2)}}$, is suppressed by the
additional factor $\alpha_s$: the behaviour of the photon-pion
transition form factor differs from that of the elastic pion form
factor where the soft-pointlike term is absent and the soft-hard terms
dominate in the large-$Q^2$ region. Representing
\begin{equation} \label{psik}
\Psi^{\mathrm{S}}_\gamma=\psi_\gamma(\vec k^2)=\frac{g_\gamma(
\vec k^2)}{\vec k^2+m^2}\ ,
\end{equation}
we determine the $\psi_\gamma(\vec k^2)$ (Fig.\ \ref{fig:wfs}b) from
data on the $\pi^0\to\gamma\gamma^*$ transition form factor
\cite{exp}. Fig.\ \ref{fig:fits}c demonstrates the description of the
data for the $\pi^0\to\gamma\gamma^*$ transition; the partial width
for the decay $\pi^0\to\gamma\gamma$, $\Gamma_{\gamma\gamma}=7.23$ eV,
provides the normalization of $\gamma\pi^0$ transition form factor
and, as result, normalises the photon wave function $\psi_\gamma(\vec
k^2)$.

To calculate $\eta\to\gamma\gamma^*$ and $\eta'\to\gamma\gamma^*$
transition form factors, we should take into account the mixing of
non-strange quark component, $n\bar n=(u\bar u+d\bar d)/\sqrt{2}$,
the strange one, $s\bar s$, and the glueball component, $G$:
\begin{eqnarray}
\Psi_\eta&=&\cos\alpha\;\left[\cos\theta\;\psi_{n\bar n}(\vec k^2)-
\sin\theta\;\psi_{s\bar s}(\vec k^2)\right]+\sin\alpha\;\psi_G\ ,
\nonumber\\
\Psi_\eta'&=&\cos\alpha'\left[\sin\theta'\;\psi_{n\bar n}(\vec k^2)+
\cos\theta'\;\psi_{s\bar s}(\vec k^2)\right]+\sin\alpha'\psi_G\ .
\nonumber\\
\label{etawfs}\end{eqnarray}
The orthogonality condition reads
\begin{equation}
\cos\alpha\cos\alpha'\ \sin(\theta'-\theta)+\sin\alpha\sin\alpha'=0\ ,
\end{equation}
determining the mixing angle $\theta'$. In the spirit of the quark
model, the universality of soft wave functions of the $0^-$ nonet is
assumed; this implies $\psi_{n\bar n}(\vec k^2)=\psi_\pi(\vec k^2)$.
For the $s\bar s$-component we take into account the
SU(3)$_{\mathrm{flavour}}$ breaking effect due to the
strange/nonstrange quark mass difference
\begin{equation}
\psi_{s\bar s}(\vec k^2)=N\frac{g_\pi(\vec k^2)}{\vec k^2
+\kappa_0^2+\Delta^2}\ ,
\end{equation}
where $\Delta^2=m_s^2-m^2$ with $m_s-m=150$ MeV. The factor $N$
corresponds to the renormalization of $\psi_{s\bar s}(\vec k^2)$ after
introducing $\Delta$.

Similarly, in the spirit of the quark model, we find for the
nonstrange and strange components of the soft photon wave function:
\begin{equation}
\psi_{\gamma\to n\bar n}(\vec k^2)=\frac{g_\gamma(\vec k^2)}{\vec
k^2+m^2}\ ,\;
\psi_{\gamma\to s\bar s}(\vec k^2)=\frac{g_\gamma(\vec k^2)}{\vec
k^2+m_s^2}.
\end{equation}
Having fixed the $n\bar n$ and $s\bar s$ components of the meson and
soft photon wave function, we calculate the $\eta\to\gamma\gamma^*$
and $\eta'\to\gamma\gamma^*$ transition form factors:
\begin{eqnarray}
F_{\eta\to\gamma\gamma^*}(Q^2)&=&\cos\alpha\;\left[\cos\theta\;
F_{n\bar n}(Q^2)-\sin\theta\;F_{s\bar s}(Q^2)\right],
\nonumber\\
F_{\eta'\to\gamma\gamma^*}(Q^2)&=&\cos\alpha'\;\left[\sin\theta'\;
F_{n\bar n}(Q^2)+\cos\theta'\;F_{s\bar s}(Q^2)\right].
\nonumber\\
\label{etaffs}\end{eqnarray}

The partial width for the transition $meson\to\gamma\gamma^*$ is equal
to
\begin{equation}
\Gamma_{\gamma\mathrm{meson}}(Q^2)=
\frac\pi 4\alpha^2m_{\mathrm{meson}}^3
F^2_{\gamma\mathrm{meson}\to\gamma\gamma*}(Q^2)\ .
\end{equation}

\begin{figure}
\centerline{\epsfig{file=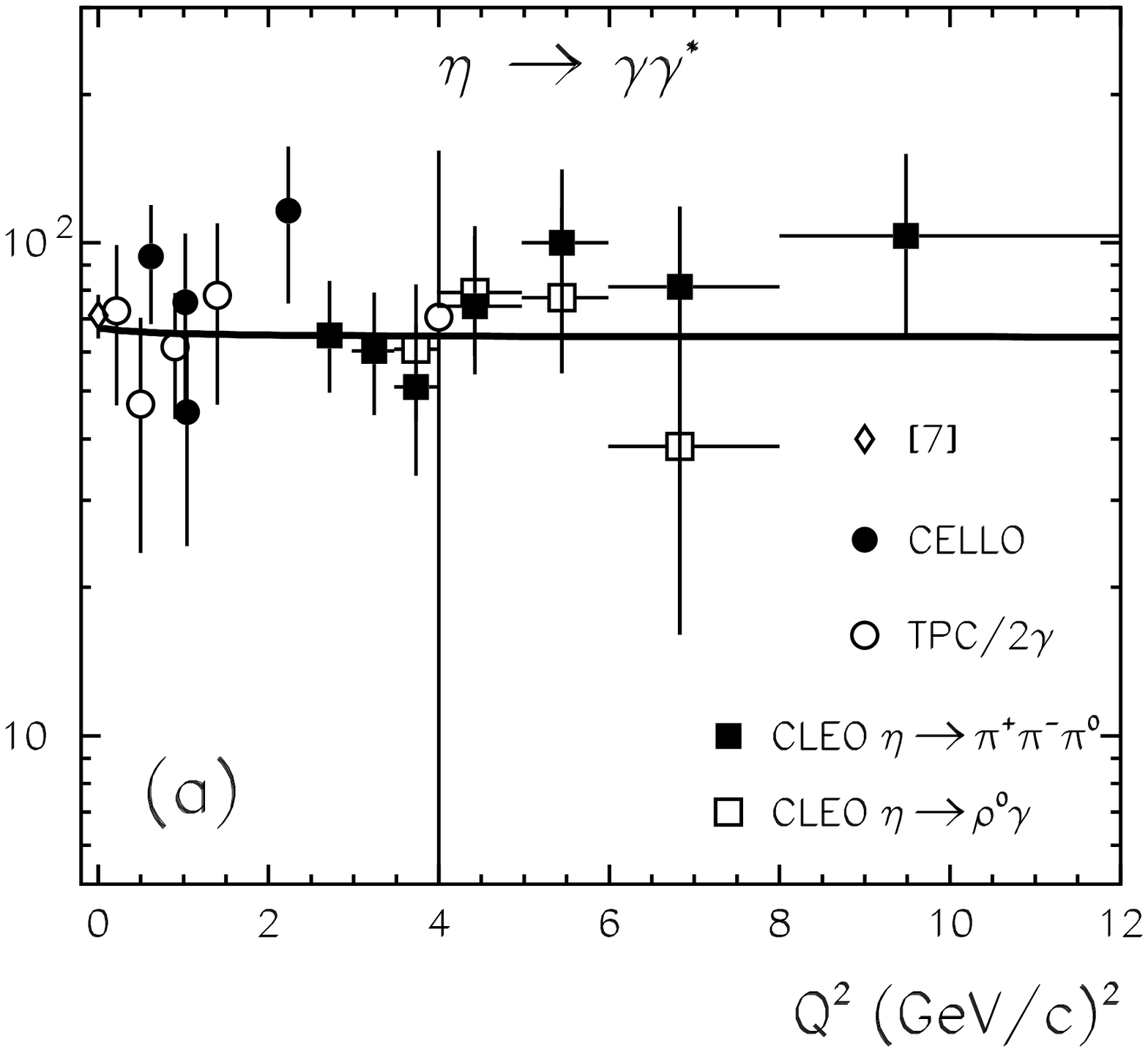, width=6cm}}
\centerline{\epsfig{file=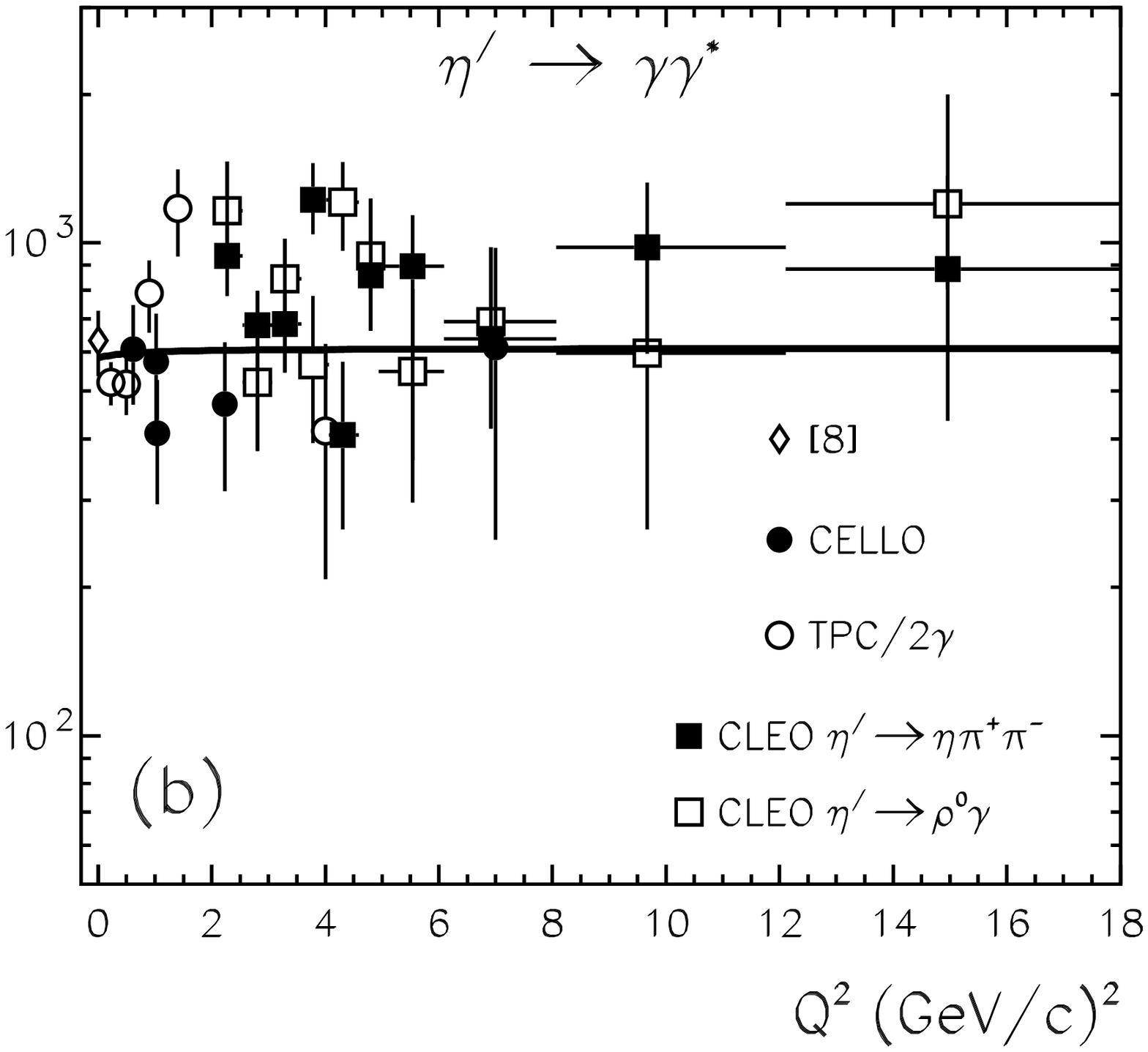,width=6cm}}
\caption{$Q^2$ dependence for the ratios (a) $
\Gamma_{\gamma\eta}(Q^2)/$ $\Gamma_{\gamma\pi^0}^{\mathrm{calc}}(Q^2)$
and (b)
$\Gamma_{\gamma\eta'}(Q^2)/\Gamma_{\gamma\pi^0}^{\mathrm{calc}}(Q^2)$
where $\Gamma_{\gamma\pi^0}^{\mathrm{calc}}(Q^2)$ is the calculated
quantity shown in Fig.\ \ref{fig:fits}c. The fitted curves correspond
to $\sin^2\alpha=0.02$, $\sin^2\alpha'=0.08$, and $\sin\theta=0.62$.}
\label{fig:eta2g} \end{figure}

Fig.\ \ref{fig:eta2g} presents the data for the partial widths
$\Gamma_{\gamma\eta}(Q^2)$ \cite{exp,gameta} and
$\Gamma_{\gamma\eta'}(Q^2)$ \cite{exp,gametap}, and their fit with
$\theta,\alpha$ and $\alpha'$ being parameters. The region of the
allowed $(\theta,\alpha,\alpha')$ values on the 90\% confidence level
is shown in Fig.\ \ref{fig:regions} (the region I).

\section{Semileptonic decays $D_s\to\eta l\nu/\eta'l\nu$}

Exclusive semileptonic decays $D_s\to\eta l\nu/\eta'l\nu$ probe the
$s\bar s$ component of $\eta$ and $\eta'$ and thus provide a test of
the mixing angle.

The decay rates are expressed through the mixing angles
$\theta,\alpha,\alpha'$ and the form factor $f_+$ of the semileptonic
transition $D_s\to\eta l\nu/\eta'l\nu$. The kinematically accessible
$q^2$-regions in $\eta$ and $\eta'$ decays are $0\le q^2\le
(M_{D_s}-M_\eta)^2$ and $0\le q^2\le (M_{D_s}-M_{\eta'})^2$.

The form factors for the semileptonic transition between pseudoscalar
mesons $P(M_1)\to P(M_2)$ induced by the quark weak transition $c\to
s$ are defined as follows \cite{iw}
\begin{equation}
\langle P(M_2,p_2)|\bar s\gamma_\mu c(0)|P(M_1,p_1)\rangle=f_
+(q^2)P_{\mu}+f_-(q^2)q_{\mu}.
\end{equation}

\begin{figure}
\centerline{\epsfig{file=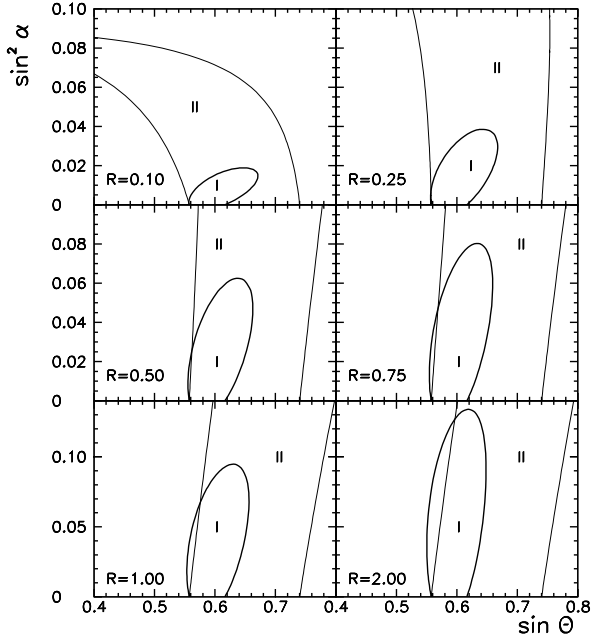,width=8cm}}
\caption{Allowed $(\theta,\alpha,\alpha')$-region: slices at
different $R=$ $\sin^2\alpha/\sin^2\alpha'$. The regions I and II are
due to the $\eta/\eta'\to\gamma\gamma^*$ and $D_s\to\eta
l\nu/\eta'l\nu$ constraints, correspondingly.} \label{fig:regions}
\end{figure}

In considering the form factor of interest we use the dispersion
relation formulation of the light-cone quark model \cite{m1}: the form
factors of the light-cone quark model of Ref.\ \cite{jaus} at
spacelike momentum transfers are represented as double spectral
representations over the invariant masses of the initial and final
$q\bar q$ pairs, and form factors in the timelike region are obtained
by analytical continuation in $q^2$.  This procedure represents the
form factors at $q^2>0$ through the light cone wave functions of the
initial and final mesons and allows direct calculation of the decay
form factors in the timelike region.  It should be emphasized that we
derive the analytical continuation in the region $q^2\le(m_2-m_1)^2$.
For the constituent quark masses used in the Isgur-Wise model
\cite{isgw2} $m_u=0.33$ GeV, $m_s=0.55$ GeV, and $m_c=1.82$ GeV which
we adopt for considering the decay process, this allows a direct
calculation of the form factor $D_s\to\eta'$ transitions in the whole
kinematical decay region $0\le q^2\le (M_{D_s}-M_\eta')^2$, as
$M_{D_s}-M_\eta'<m_c-m_s$. For the $D_s\to \eta$ transition this is
not the case as $M_{D_s}-M_{\eta}>m_c-m_s$.  For this mode we directly
calculate the form factors in the region $0\le q^2\le (m_2-m_1)^2$ and
perform numerical extrapolation in $(m_c-m_s)^2\le q^2\le
M_{D_s}-M_{\eta}$. All relevant technical details can be found in
\cite{m1}.

The transition form factors are expressed through the light-cone
wave functions of the initial and final mesons. For $\eta$ and $\eta'$
we know such functions from the previous consideration. For the $D_s$
meson we assume the $D_s$ wave function to be approximated by a simple
one-parameter exponential function $w(k)=\exp(-k^2/2\beta^2)$ and
adopt the value $\beta=0.56$ GeV from the Isgur-Wise model
\cite{isgw2}.

The results of calculating the form factors in the region
$q^2<(m_c-m_s)^2$ are fitted by the function
\begin{equation}
f_+(q^2)=f_+(0)/[1-\alpha_1q^2+\alpha_2q^4]
\end{equation}
with better than $0.5\%$ accuracy. We found $f_+(0)=0.8$, $\alpha_1
=0.192\mbox{ GeV}^{-2}$, and $\alpha_2=0.008\mbox{ GeV}^{-2}$.

For the form factor $f_+$ in the region $(m_c-m_s)^2\le q^2\le
M_{D_s}-M_{\eta}$ this formula is used for numerical extrapolation.
Numerical analysis shows the accuracy of this extrapolation procedure
to be very high. The decay rates are calculated from the form factors
via the formulae from \cite{gs}. The calculated decay rates depend on
the content of $\eta$ and $\eta'$ mesons, and their ratio with
$|V_{cs}|=0.975$ reads
\begin{equation}
\frac{\Gamma(D_s\to\eta'l\nu)}{\Gamma(D_s\to\eta l\nu)}= 0.28
\frac{\cos^2\theta'\;\cos^2\alpha'}{\sin^2\theta\;\cos^2\alpha}\ .
\end{equation}

It should be pointed out that the obtained decay rates are calculated
neglecting a nontrivial structure of the constituent quark transition
form factor. This is a conventional but rather crude approximation:
in particular, the quark transition form factor should contain a pole
at $q^2=M_{\mathrm{res}}^2$ with $M_{\mathrm{res}}$ the mass of a
resonance with appropriate quantum numbers. It is not clear yet
whether the transition form factor differs significantly from unity or
not in the kinematical decay region. Anyway, the transition form
factor is a rising function at $q^2>0$. This property yields an
important consequence for the ratio of the decay rates
$\Gamma(D_s\to\eta')/\Gamma(D_s\to\eta)$: as the phase space of the
decay $D_s\to\eta$ is larger than that of the decay $D_s\to \eta'$,
taking into account the (rising) constituent quark form factor will
decrease the theoretical value of the ratio. We can try to estimate
the effect caused by a nontrivial form factor
taking a simple monopole $q^2$ dependence:
\begin{equation}
f_c(q^2)=\frac 1{1-q^2/M_{\mathrm{res}}^2}
\end{equation}
with $M_{\mathrm{res}}=M_{D^*_{s}}=2.1$ GeV. This yields the following
shift in the predicted ratio
\begin{equation}
\frac{\Gamma(D_s\to\eta'l\nu)}{\Gamma(D_s\to\eta l\nu)}= 0.23
\frac{\cos^2\theta'\;\cos^2\alpha'}{\sin^2\theta\;\cos^2\alpha}\ .
\end{equation}

CLEO gives $\Gamma(D_s\to\eta')/\Gamma(D_s\to\eta)=0.35\pm0.16$
\cite{dsexp} implying a new restriction on the region of the allowed
$(\theta,\alpha,\alpha')$: the region II on Fig.\ \ref{fig:regions}.

\section{Conclusion}

Based on the data for transitions $\eta/\eta'\to\gamma\gamma*$
\cite{exp,gameta,gametap} and $D_s\to\eta l\nu/\eta'l\nu$
\cite{dsexp}, we have determined the region of the allowed mixing
angles for $\eta_1,\eta_8$ and glueball component in $\eta$ and
$\eta'$. The determined $\eta_1/\eta_8$ mixing angle
$\theta_{\mathrm{P}}=-17.0^\circ\pm2.6^\circ$ is in between the values
given by the linear and quadratic mass formulae. It looks reasonable
to suppose the glueball component in $\eta'$ meson prevails that in
$\eta$ meson, $R=\sin^2\alpha/\sin^2\alpha'<1$: in this case the
glueball component probabilities are in the borders
$\sin^2\alpha\le0.08$, $\sin^2\alpha'\le0.12$. The significant values
of the glueball components in $\eta$ and $\eta'$ are in an agreement
with enlarged production of these mesons in the radiative $J/\psi$
decay \cite{datagr}. We think that the decays
$J/\psi\to\gamma\eta/\gamma\eta'$ can provide an additional important
information about glueball components in $\eta$ and $\eta'$. However
these decays need a special consideration that is beyond our present
analysis.

\section*{Acknowledgement}
We are indebted to F.E. Close, S.S. Gershtein, and B.S. Zou for
valuable remarks and discussions. VVA thanks F.E. Close for
hospitality at Ruther\-ford--Appleton Laboratory when carrying out
this work. This investigation is supported by INTAS-RFBR grant
95-0267.

\end{document}